\documentclass[aip,jcp,reprint,noshowkeys,superscriptaddress]{revtex4-1}
\usepackage{graphicx,dcolumn,bm,xcolor,microtype,multirow,amscd,amsmath,amssymb,amsfonts,physics,longtable,wrapfig,bbold,siunitx,xspace}
\usepackage[version=4]{mhchem}

\usepackage[utf8]{inputenc}
\usepackage[T1]{fontenc}

\usepackage{hyperref}
\hypersetup{
    colorlinks,
    linkcolor={red!50!black},
    citecolor={red!70!black},
    urlcolor={red!80!black}
}

\usepackage{listings}
\definecolor{codegreen}{rgb}{0.58,0.4,0.2}
\definecolor{codegray}{rgb}{0.5,0.5,0.5}
\definecolor{codepurple}{rgb}{0.25,0.35,0.55}
\definecolor{codeblue}{rgb}{0.30,0.60,0.8}
\definecolor{backcolour}{rgb}{0.98,0.98,0.98}
\definecolor{mygray}{rgb}{0.5,0.5,0.5}

\definecolor{sqred}{rgb}{0.85,0.1,0.1}
\definecolor{sqgreen}{rgb}{0.25,0.65,0.15}
\definecolor{sqorange}{rgb}{0.90,0.50,0.15}
\definecolor{sqblue}{rgb}{0.10,0.3,0.60}

\lstdefinestyle{mystyle}{
    backgroundcolor=\color{backcolour},
    commentstyle=\color{codegreen},
    keywordstyle=\color{codeblue},
    numberstyle=\tiny\color{codegray},
    stringstyle=\color{codepurple},
    basicstyle=\ttfamily\footnotesize,
    breakatwhitespace=false,
    breaklines=true,
    captionpos=b,
    keepspaces=true,
    numbers=left,
    numbersep=5pt,
    numberstyle=\ttfamily\tiny\color{mygray},
    showspaces=false,
    showstringspaces=false,
    showtabs=false,
    tabsize=2
  }

  \newcolumntype{d}{D{.}{.}{-1}}

\newcommand{\SupInf}{\textcolor{blue}{Supporting Information}}

\DeclareMathOperator{\sgn}{sgn}

\newcommand{\LCPQ}{Laboratoire de Chimie et Physique Quantiques (UMR 5626), Universit\'e de Toulouse, CNRS, UPS, France}

\begin{document}	

\title{A similarity renormalization group approach to Green's function methods}

\author{Antoine \surname{Marie}}
	\email{amarie@irsamc.ups-tlse.fr}
	\affiliation{\LCPQ}

\author{Pierre-Fran\c{c}ois \surname{Loos}}
	\email{loos@irsamc.ups-tlse.fr}
	\affiliation{\LCPQ}

\begin{abstract}
The family of Green's function methods based on the $GW$ approximation has gained popularity in the electronic structure theory thanks to its accuracy in weakly correlated systems combined with its cost-effectiveness.
Despite this, self-consistent versions still pose challenges in terms of convergence.
A recent study \href{https://doi.org/10.1063/5.0089317}{[J. Chem. Phys. 156, 231101 (2022)]} has linked these convergence issues to the intruder-state problem.
In this work, a perturbative analysis of the similarity renormalization group (SRG) approach is performed on Green's function methods.
The SRG formalism enables us to derive, from first principles, the expression of a naturally static and Hermitian form of the self-energy that can be employed in quasiparticle self-consistent $GW$ (qs$GW$) calculations.
The resulting SRG-based regularized self-energy significantly accelerates the convergence of qs$GW$ calculations, slightly improves the overall accuracy, and is straightforward to implement in existing code.
\bigskip
\begin{center}
	\boxed{\includegraphics[width=0.5\linewidth]{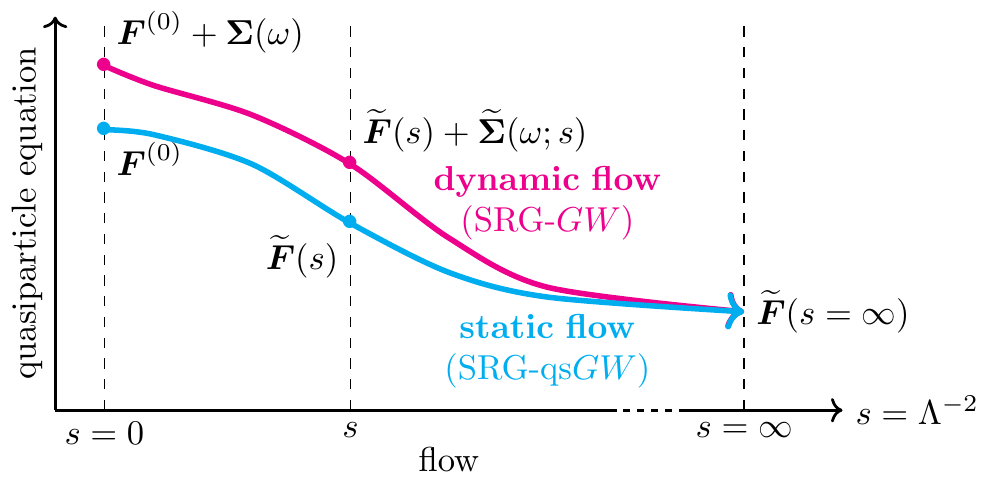}}
\end{center}
\bigskip
\end{abstract}

\maketitle

\section{Introduction}
\label{sec:intro}

The one-body Green's function provides a natural and elegant way to access the charged excitation energies of a physical system. \cite{CsanakBook,FetterBook,Martin_2016,Golze_2019}
The non-linear Hedin equations consist of a closed set of equations leading to the exact interacting one-body Green's function and, therefore, to a wealth of properties such as the total energy, density, ionization potentials, electron affinities, as well as spectral functions, without the explicit knowledge of the wave functions associated with the neutral and charged electronic states of the system. \cite{Hedin_1965}
Unfortunately, solving exactly Hedin's equations is usually out of reach and one must resort to approximations.
In particular, the $GW$ approximation, \cite{Hedin_1965,Aryasetiawan_1998,Onida_2002,Reining_2017,Golze_2019,Bruneval_2021} which has been first introduced in the context of solids \cite{Strinati_1980,Strinati_1982a,Strinati_1982b,Hybertsen_1985,Hybertsen_1986,Godby_1986,Godby_1987,Godby_1987a,Godby_1988,Blase_1995} and is now widely applied to molecular systems, \cite{Rohlfing_1999a,Horst_1999,Puschnig_2002,Tiago_2003,Rocca_2010,Boulanger_2014,Jacquemin_2015a,Bruneval_2015,Jacquemin_2015b,Hirose_2015,Jacquemin_2017a,Jacquemin_2017b,Rangel_2017,Krause_2017,Gui_2018,Blase_2018,Liu_2020,Li_2017,Li_2019,Li_2020,Li_2021,Blase_2020,Holzer_2018a,Holzer_2018b,Loos_2020e,Loos_2021,McKeon_2022} yields accurate charged excitation energies for weakly correlated systems \cite{Hung_2017,vanSetten_2015,vanSetten_2018,Caruso_2016,Korbel_2014,Bruneval_2021} at a relatively low computational cost. \cite{Foerster_2011,Liu_2016,Wilhelm_2018,Forster_2021,Duchemin_2019,Duchemin_2020,Duchemin_2021}

The $GW$ method approximates the self-energy $\Sigma$ which relates the exact interacting Green's function $G$ to a non-interacting reference version $G_0$ through a Dyson equation of the form
\begin{equation}
  \label{eq:dyson}
  G(1,2) = G_0(1,2) + \int d(34) G_0(1,3)\Sigma(3,4) G(4,2),
\end{equation}
where $1 = (\bx_1, t_1)$ is a composite coordinate gathering spin-space and time variables.
The self-energy encapsulates all the Hartree-exchange-correlation effects which are not taken into account in the reference system.
Approximating $\Sigma$ as the first-order term of its perturbative expansion with respect to the screened Coulomb potential $W$ yields the so-called $GW$ approximation \cite{Hedin_1965,Martin_2016}
\begin{equation}
  \label{eq:gw_selfenergy}
  \Sigma(1,2) = \ii G(1,2) W(1,2).
\end{equation}
Diagrammatically, $GW$ involves a resummation of the (time-dependent) direct ring diagrams via the computation of the random-phase approximation (RPA) polarizability \cite{Ren_2012,Chen_2017} and is thus particularly well suited for weak correlation.

Despite a wide range of successes, many-body perturbation theory has well-documented limitations. \cite{Kozik_2014,Stan_2015,Rossi_2015,Tarantino_2017,Schaefer_2013,Schaefer_2016,Gunnarsson_2017,vanSetten_2015,Maggio_2017a,Duchemin_2020}
For example, modeling core-electron spectroscopy requires core ionization energies which have been proven to be challenging for routine $GW$ calculations. \cite{vanSetten_2018,Golze_2018,Golze_2020,Li_2022}
Many-body perturbation theory can also be used to access optical excitation energies through the Bethe-Salpeter equation. \cite{Salpeter_1951,Strinati_1988,Blase_2018,Blase_2020} However, the accuracy is not yet satisfying for triplet excited states, where instabilities often occur. \cite{Bruneval_2015,Jacquemin_2017a,Jacquemin_2017b,Holzer_2018a}
Therefore, even if $GW$ offers a good trade-off between accuracy and computational cost, some situations might require higher precision.
Unfortunately, defining a systematic way to go beyond $GW$ via the inclusion of vertex corrections has been demonstrated to be a tricky task. \cite{Baym_1961,Baym_1962,DeDominicis_1964a,DeDominicis_1964b,Bickers_1989a,Bickers_1989b,Bickers_1991,Hedin_1999,Bickers_2004,Shirley_1996,DelSol_1994,Schindlmayr_1998,Morris_2007,Shishkin_2007b,Romaniello_2009a,Romaniello_2012,Gruneis_2014,Hung_2017,Maggio_2017b,Mejuto-Zaera_2022}
For example, Lewis and Berkelbach have shown that naive vertex corrections can even worsen the quasiparticle energies with respect to $GW$. \cite{Lewis_2019}
We refer the reader to the recent review by Golze and co-workers \cite{Golze_2019} for an extensive list of current challenges in Green's function methods. 

Many-body perturbation theory also suffers from the infamous intruder-state problem,\cite{Andersson_1994,Andersson_1995a,Roos_1995,Forsberg_1997,Olsen_2000,Choe_2001} where they manifest themselves as solutions of the quasiparticle equation with non-negligible spectral weights.
In some cases, this transfer of spectral weight makes it difficult to distinguish between a quasiparticle and a satellite.
These multiple solutions hinder the convergence of partially self-consistent schemes, \cite{Veril_2018,Forster_2021,Monino_2022} such as eigenvalue-only self-consistent $GW$ \cite{Shishkin_2007a,Blase_2011b,Marom_2012,Kaplan_2016,Wilhelm_2016} (ev$GW$) and quasiparticle self-consistent $GW$ \cite{Faleev_2004,vanSchilfgaarde_2006,Kotani_2007,Ke_2011,Kaplan_2016} (qs$GW$). 
The simpler one-shot $G_0W_0$ scheme \cite{Strinati_1980,Hybertsen_1985a,Hybertsen_1986,Godby_1988,Linden_1988,Northrup_1991,Blase_1994,Rohlfing_1995,Shishkin_2007a} is also impacted by these intruder states, leading to discontinuities and/or irregularities in a variety of physical quantities including charged and neutral excitation energies, correlation and total energies.\cite{Loos_2018b,Veril_2018,Loos_2020e,Berger_2021,DiSabatino_2021,Monino_2022,Scott_2023}
These convergence problems and discontinuities can even happen in the weakly correlated regime where the $GW$ approximation is supposed to be valid.

In a recent study, Monino and Loos showed that the discontinuities could be removed by the introduction, in the quasiparticle equation, of a regularizer inspired by the similarity renormalization group (SRG). \cite{Monino_2022}
Encouraged by this study and the recent successes of regularization schemes in many-body quantum chemistry methods, such as in single- and multi-reference perturbation theory, \cite{Lee_2018a,Shee_2021,Evangelista_2014b,ChenyangLi_2019a,Battaglia_2022,Coveney_2023} the present work investigates the application of the SRG formalism in $GW$-based methods.
In particular, we focus here on the possibility of curing the qs$GW$ convergence issues using the SRG.

The SRG formalism has been developed independently by Wegner \cite{Wegner_1994} in the context of condensed matter systems and Glazek \& Wilson \cite{Glazek_1993,Glazek_1994} in light-front quantum field theory.
This formalism has been introduced in quantum chemistry by White \cite{White_2002} before being explored in more detail by Evangelista and coworkers in the context of multi-reference electron correlation theories. \cite{Evangelista_2014b,ChenyangLi_2015, ChenyangLi_2016,ChenyangLi_2017,ChenyangLi_2018,ChenyangLi_2019a,Zhang_2019,ChenyangLi_2021,Wang_2021,Wang_2023}
The SRG has also been successful in the context of nuclear structure theory, where it was first developed as a mature computational tool thanks to the work of several research groups.
\cite{Bogner_2007,Tsukiyama_2011,Tsukiyama_2012,Hergert_2013,Hergert_2016,Frosini_2022,Frosini_2022a,Frosini_2022b} 
See Ref.~\onlinecite{Hergert_2016} for a recent review in this field.

The SRG transformation aims at decoupling an internal (or reference) space from an external space while incorporating information about their coupling in the reference space.
This process often results in the appearance of intruder states. \cite{Evangelista_2014b,ChenyangLi_2019a}
However, SRG is particularly well-suited to avoid these because the decoupling of each external configuration is inversely proportional to its energy difference with the reference space.
By definition, intruder states have energies that are close to the reference energy, and, therefore, are the last to be decoupled.
By stopping the SRG transformation once all external configurations except the intruder states have been decoupled,
correlation effects between the internal and external spaces can be incorporated (or folded) without the presence of intruder states.

The goal of this manuscript is to determine if the SRG formalism can effectively address the issue of intruder states in many-body perturbation theory, as it has in other areas of electronic and nuclear structure theory.
This open question will lead us to an intruder-state-free static approximation of the self-energy derived from first-principles that can be employed in partially self-consistent $GW$ calculations.
Note that throughout the manuscript we focus on the $GW$ approximation but the subsequent derivations can be straightforwardly applied to other self-energies such as the one derived from second-order Green's function \cite{Casida_1989,Casida_1991,SzaboBook,Stefanucci_2013,Ortiz_2013,Phillips_2014,Phillips_2015,Rusakov_2014,Rusakov_2016,Hirata_2015,Hirata_2017,Backhouse_2021,Backhouse_2020b,Backhouse_2020a,Pokhilko_2021a,Pokhilko_2021b,Pokhilko_2022} or the $T$-matrix approximation.\cite{Liebsch_1981,Bickers_1989a,Bickers_1991,Katsnelson_1999,Katsnelson_2002,Zhukov_2005,vonFriesen_2010,Romaniello_2012,Gukelberger_2015,Muller_2019,Friedrich_2019,Biswas_2021,Zhang_2017,Li_2021b,Loos_2022}

The manuscript is organized as follows.
We begin by reviewing the $GW$ approximation in Sec.~\ref{sec:gw} and then briefly introduce the SRG formalism in Sec.~\ref{sec:srg}.
A perturbative analysis of SRG applied to $GW$ is presented in Sec.~\ref{sec:srggw}.
The computational details are provided in Sec.~\ref{sec:comp_det} before turning to the results (Sec.~\ref{sec:results}).
Our conclusions are drawn in Sec.~\ref{sec:conclusion}.
Unless otherwise stated, atomic units are used throughout.


\section{The $GW$ approximation}
\label{sec:gw}

The central equation of many-body perturbation theory based on Hedin's equations is the so-called dynamical and non-Hermitian quasiparticle equation which, within the $GW$ approximation, reads
\begin{equation}
  \label{eq:quasipart_eq}
  \qty[ \bF + \bSig(\omega = \epsilon_p) ] \psi_p(\bx) = \epsilon_p \psi_p(\bx),
\end{equation}
where $\bF$ is the Fock matrix in the orbital basis \cite{SzaboBook} and $\bSig(\omega)$ is (the correlation part of) the $GW$ self-energy. 
Both are $K \times K$ matrices with $K$ the number of one-electron orbitals.
Throughout the manuscript, the indices $p,q,r,s$ are general orbitals while $i,j,k,l$ and $a,b,c,d$ refer to occupied and virtual orbitals, respectively.
The indices $\mu$ and $\nu$ are composite indices, that is, $\nu=(ia)$, referring to neutral (single) excitations.

The self-energy can be physically understood as a correction to the Hartree-Fock (HF) problem (represented by $\bF$) accounting for dynamical screening effects.
Similarly to the HF case, Eq.~\eqref{eq:quasipart_eq} has to be solved self-consistently but the dynamical and non-Hermitian nature of $\bSig(\omega)$, as well as its functional form, makes it much more challenging to solve from a practical point of view.

The matrix elements of $\bSig(\omega)$ have the following closed-form expression \cite{Hedin_1999,Tiago_2006,Bruneval_2012,vanSetten_2013,Bruneval_2016}
\begin{equation}
  \label{eq:GW_selfenergy}
  \Sigma_{pq}(\omega) 
  = \sum_{i\nu} \frac{W_{pi}^{\nu} W_{qi}^{\nu}}{\omega - \epsilon_i + \Omega_{\nu} - \ii \eta} 
  + \sum_{a\nu} \frac{W_{pa}^{\nu} W_{qa}^{\nu}}{\omega - \epsilon_a - \Omega_{\nu} + \ii \eta},
\end{equation}
where $\eta$ is a positive infinitesimal and the screened two-electron integrals are
\begin{equation}
  \label{eq:GW_sERI}
  W_{pq}^{\nu} = \sum_{ia}\eri{pi}{qa}\qty(\bX+\bY)_{ia}^{\nu},
\end{equation}
with $\bX$ and $\bY$ the components of the eigenvectors of the direct (\ie without exchange) RPA problem defined as 
\begin{equation}
  \label{eq:full_dRPA}
  \mqty( \bA & \bB \\ -\bB & -\bA ) \mqty( \bX & \bY \\ \bY & \bX ) =   \mqty( \bX & \bY \\ \bY & \bX ) \mqty( \boldsymbol{\Omega} & \bO \\ \bO & -\boldsymbol{\Omega} ),
\end{equation}
with
\begin{subequations}
\begin{align}
  A_{ia,jb} & = (\epsilon_a - \epsilon_i) \delta_{ij}\delta_{ab} + \eri{ib}{aj},
  \\
  B_{ia,jb} & = \eri{ij}{ab},
\end{align}
\end{subequations}
and where
\begin{equation}
	\braket{pq}{rs} = \iint \frac{\SO{p}(\bx_1) \SO{q}(\bx_2)\SO{r}(\bx_1) \SO{s}(\bx_2)  }{\abs{\br_1 - \br_2}} d\bx_1 d\bx_2
\end{equation}
are bare two-electron integrals in the spin-orbital basis.

The diagonal matrix $\boldsymbol{\Omega}$ contains the positive eigenvalues of the RPA problem defined in Eq.~\eqref{eq:full_dRPA} and its elements $\Omega_\nu$ appear in Eq.~\eqref{eq:GW_selfenergy}.

As mentioned above, because of the frequency dependence of the self-energy, solving exactly the quasiparticle equation \eqref{eq:quasipart_eq} is a rather complicated task. 
Hence, several approximate schemes have been developed to bypass full self-consistency.
The most popular strategy is the one-shot (perturbative) $GW$ scheme, $G_0W_0$, where the self-consistency is completely abandoned, and the off-diagonal elements of Eq.~\eqref{eq:quasipart_eq} are neglected.
Assuming a HF starting point, this results in $K$ quasiparticle equations that read
\begin{equation}
  \label{eq:G0W0}
  \epsilon_p^{\HF} + \Sigma_{pp}(\omega) - \omega = 0,
\end{equation}
where $\Sigma_{pp}(\omega)$ are the diagonal elements of $\bSig$ and $\epsilon_p^{\HF}$ are the HF orbital energies.
The previous equations are non-linear with respect to $\omega$ and therefore have multiple solutions $\epsilon_{p,z}$ for a given $p$ (where the index $z$ is numbering solutions).
These solutions can be characterized by their spectral weight given by the renormalization factor
\begin{equation}
  \label{eq:renorm_factor}
  0 \leq Z_{p,z} = \qty[ 1 - \eval{\pdv{\Sigma_{pp}(\omega)}{\omega}}_{\omega=\epsilon_{p,z}} ]^{-1} \leq 1.
\end{equation}
The solution with the largest weight $Z_p \equiv Z_{p,z=0}$ is referred to as the quasiparticle while the others are known as satellites (or shake-up transitions).
However, in some cases, Eq.~\eqref{eq:G0W0} can have two (or more) solutions with similar weights, hence the quasiparticle is not well-defined.

One obvious drawback of the one-shot scheme mentioned above is its starting-point dependence.
Indeed, in Eq.~\eqref{eq:G0W0} we choose to rely on HF orbital energies but this is arbitrary and one could have chosen Kohn-Sham energies (and orbitals) instead.
As commonly done, one can even ``tune'' the starting point to obtain the best possible one-shot $GW$ quasiparticle energies. \cite{Korzdorfer_2012,Marom_2012,Bruneval_2013,Gallandi_2015,Caruso_2016,Gallandi_2016}

Alternatively, one may solve iteratively the set of quasiparticle equations \eqref{eq:G0W0} to reach convergence of the quasiparticle energies, leading to the partially self-consistent scheme named ev$GW$.
However, if one of the quasiparticle equations does not have a well-defined quasiparticle solution, reaching self-consistency can be challenging, if not impossible. 
Even at convergence, the starting point dependence is not totally removed as the quasiparticle energies still depend on the initial set of orbitals. \cite{Marom_2012}

In order to update both the orbitals and their corresponding energies, one must consider the off-diagonal elements in $\bSig(\omega)$.
To avoid solving the non-Hermitian and dynamic quasiparticle equation defined in Eq.~\eqref{eq:quasipart_eq}, one can resort to the qs$GW$ scheme in which $\bSig(\omega)$ is replaced by a static approximation $\bSig^{\qsGW}$.
Then, the qs$GW$ equations are solved via a standard self-consistent field procedure similar to the HF algorithm where $\bF$ is replaced by $\bF + \bSig^{\qsGW}$.
Various choices for $\bSig^{\qsGW}$ are possible but the most popular is the following Hermitian approximation
\begin{equation}
  \label{eq:sym_qsgw}
  \Sigma_{pq}^{\qsGW} = \frac{1}{2}\Re[\Sigma_{pq}(\epsilon_p) + \Sigma_{pq}(\epsilon_q) ],
\end{equation}
which was first introduced by Faleev and co-workers \cite{Faleev_2004,vanSchilfgaarde_2006,Kotani_2007,Lei_2022} before being derived by Ismail-Beigi as the effective Hamiltonian that minimizes the length of the gradient of the Klein functional for non-interacting Green's functions. \cite{Ismail-Beigi_2017}
The corresponding matrix elements are
\begin{equation}
  \label{eq:sym_qsGW}
  \Sigma_{pq}^{\qsGW} 
  = \frac{1}{2} \sum_{r\nu}  \qty[ \frac{\Delta_{pr}^{\nu}}{(\Delta_{pr}^{\nu})^2 
  + \eta^2} +\frac{\Delta_{qr}^{\nu}}{(\Delta_{qr}^{\nu})^2 + \eta^2} ] W_ {pr}^{\nu} W_{qr}^{\nu},
\end{equation}
with $\Delta_{pr}^{\nu} = \epsilon_p - \epsilon_r - \sgn(\epsilon_r-\epsilon_F)\Omega_\nu$ (where $\epsilon_F$ is the energy of the Fermi level).
One of the main results of the present manuscript is the derivation, from first principles, of an alternative static Hermitian form for the qs$GW$ self-energy. 

Once again, in cases where multiple solutions have large spectral weights, self-consistency can be difficult to reach at the qs$GW$ level.
Multiple solutions of Eq.~\eqref{eq:G0W0} arise due to the $\omega$ dependence of the self-energy.
Therefore, by suppressing this dependence, the static approximation relies on the fact that there is well-defined quasiparticle solutions.
If it is not the case, the self-consistent qs$GW$ scheme inevitably oscillates between solutions with large spectral weights. \cite{Forster_2021}

The satellites causing convergence issues are the above-mentioned intruder states. \cite{Monino_2022}
One can deal with them by introducing \textit{ad hoc} regularizers.
For example, the $\ii\eta$ term in the denominators of Eq.~\eqref{eq:GW_selfenergy}, sometimes referred to as a broadening parameter linked to the width of the quasiparticle peak, is similar to the usual imaginary-shift regularizer employed in various other theories plagued by the intruder-state problem. \cite{Surjan_1996,Forsberg_1997,Monino_2022,Battaglia_2022}.

However, this $\eta$ parameter is required to define the Fourier transformation between time and energy representation and should theoretically be set to zero. \cite{Martin_2016}
Several other regularizers are possible \cite{Stuck_2013,Rostam_2017,Lee_2018a,Evangelista_2014b,Shee_2021,Coveney_2023} and, in particular, it was shown in Ref.~\onlinecite{Monino_2022} that a regularizer inspired by the SRG had some advantages over the imaginary shift.
Nonetheless, it would be more rigorous, and more instructive, to obtain this regularizer from first principles by applying the SRG formalism to many-body perturbation theory.
This is one of the aims of the present work.

\section{The similarity renormalization group}
\label{sec:srg}

The SRG method aims at continuously transforming a general Hamiltonian matrix to its diagonal form, or more often, to a block-diagonal form.
Hence, the first step is to decompose this Hamiltonian matrix 
\begin{equation}
  \bH = \bH^\text{d} + \bH^\text{od},
\end{equation}
into an off-diagonal part, $\bH^\text{od}$, that we aim at removing and the remaining diagonal part, $\bH^\text{d}$.

This transformation can be performed continuously via a unitary matrix $\bU(s)$, as follows:
\begin{equation}
  \label{eq:SRG_Ham}
  \bH(s) = \bU(s) \, \bH \, \bU^\dag(s),
\end{equation}
where the flow parameter $s$ controls the extent of the decoupling and is related to an energy cutoff $\Lambda=s^{-1/2}$.
For a given value of $s$, only states with energy difference (with respect to the reference space) greater than $\Lambda$ are decoupled from the reference space, hence avoiding potential intruders.
By definition, the boundary conditions are $\bH(s=0) = \bH$ [or $\bU(s=0) = \bI$] and $\bH^\text{od}(s=\infty) = \bO$.

An evolution equation for $\bH(s)$ can be easily obtained by differentiating Eq.~\eqref{eq:SRG_Ham} with respect to $s$, yielding the flow equation
\begin{equation}
  \label{eq:flowEquation}
  \dv{\bH(s)}{s} = \comm{\boldsymbol{\eta}(s)}{\bH(s)},
\end{equation}
where $\boldsymbol{\eta}(s)$, the flow generator, is defined as
\begin{equation}
  \boldsymbol{\eta}(s) = \dv{\bU(s)}{s}	\bU^\dag(s)	= - \boldsymbol{\eta}^\dag(s).
\end{equation}
The flow equation can then be approximately solved by introducing an approximate form of $\boldsymbol{\eta}(s)$.

In this work, we consider Wegner's canonical generator \cite{Wegner_1994}
\begin{equation}
  \boldsymbol{\eta}^\text{W}(s) = \comm{\bH^\text{d}(s)}{\bH(s)} = \comm{\bH^\text{d}(s)}{\bH^\text{od}(s)},
\end{equation}
which satisfies the following condition \cite{Kehrein_2006}
\begin{equation}
  \label{eq:derivative_trace}
  \dv{s}\text{Tr}\left[  \bH^\text{od}(s)^\dagger \bH^\text{od}(s) \right] \leq 0.
\end{equation}
This implies that the matrix elements of the off-diagonal part decrease in a monotonic way throughout the transformation.
Moreover, the coupling coefficients associated with the highest-energy determinants are removed first as we shall evidence in the perturbative analysis below.
The main drawback of this generator is that it generates a stiff set of ODE which is therefore difficult to solve numerically. 
However, here we will not tackle the full SRG problem but only consider analytical low-order perturbative expressions. 
Hence, we will not be affected by this problem. \cite{Evangelista_2014,Hergert_2016}

Let us now perform the perturbative analysis of the SRG equations.
For $s=0$, the initial problem is
\begin{equation}
  \bH(0) = \bH^\text{d}(0) + \lambda \bH^\text{od}(0),
\end{equation}
where $\lambda$ is the usual perturbation parameter and the off-diagonal part of the Hamiltonian has been defined as the perturbation.
For finite values of $s$, we have the following perturbation expansion of the Hamiltonian
\begin{equation}
  \label{eq:perturbation_expansionH}
  \bH(s) = \bH^{(0)}(s) + \lambda ~ \bH^{(1)}(s) + \lambda^2 \bH^{(2)}(s) + \cdots.
\end{equation}
The generator $\boldsymbol{\eta}(s)$ admits a similar perturbation expansion.
Then, as performed in Sec.~\ref{sec:srggw}, one can collect order by order the terms in Eq.~\eqref{eq:flowEquation} and solve analytically the low-order differential equations.

\section{Regularized $GW$ approximation}
\label{sec:srggw}

Here, we combine the concepts of the two previous subsections and apply the SRG method to the $GW$ formalism.
However, to do so, one must identify the coupling terms in Eq.~\eqref{eq:quasipart_eq}, which is not straightforward.
A way around this problem is to transform Eq.~\eqref{eq:quasipart_eq} to an equivalent upfolded form which elegantly highlights the coupling terms.
Indeed, the $GW$ quasiparticle equation is equivalent to the diagonalization of the following matrix \cite{Bintrim_2021,Tolle_2023}
\begin{equation}
  \label{eq:GWlin}
  \begin{pmatrix}
    \bF                     &\bW^{\text{2h1p}} &	\bW^{\text{2p1h}}	\\
    (\bW^{\text{2h1p}})^\dag  &\bC^{\text{2h1p}}  &	\bO		\\
    (\bW^{\text{2p1h}})^\dag  &\bO              &	\bC^{\text{2p1h}}	\\
  \end{pmatrix},
\end{equation}
where the 2h1p and 2p1h matrix elements are
\begin{subequations}
  \begin{align}
    C^\text{2h1p}_{i\nu,j\mu} & =  \left(\epsilon_i - \Omega_\nu\right)\delta_{ij}\delta_{\nu\mu},
    \\
    C^\text{2p1h}_{a\nu,b\mu} & = \left(\epsilon_a + \Omega_\nu\right)\delta_{ab}\delta_{\nu\mu},
  \end{align}
\end{subequations}
and the corresponding coupling blocks read [see Eq.~(\ref{eq:GW_sERI})]
\begin{align}
  W^\text{2h1p}_{p,i\nu} & = W_{pi}^{\nu},
  &
    W^\text{2p1h}_{p,a\nu} & = W_{pa}^{\nu}.
\end{align}

The usual $GW$ non-linear equation can be obtained by applying L\"owdin partitioning technique \cite{Lowdin_1963} to Eq.~\eqref{eq:GWlin} yielding \cite{Bintrim_2021} 
\begin{equation}
  \begin{split}
    \label{eq:downfolded_sigma}
  \bSig(\omega)
  & = \bW^{\hhp} \qty(\omega \bI - \bC^{\hhp})^{-1}  (\bW^{\hhp})^\dag
  \\
  & + \bW^{\pph} \qty(\omega \bI - \bC^{\pph})^{-1}  (\bW^{\pph})^\dag, 
\end{split}
\end{equation}
which can be further developed to recover exactly Eq.~\eqref{eq:GW_selfenergy}.

Equations \eqref{eq:GWlin} and \eqref{eq:quasipart_eq} yield exactly the same quasiparticle and satellite energies but one is linear and the other is not.
The price to pay for this linearity is that the size of the matrix in the former is $\order{K^3}$ while it is only $\order{K}$ in the latter.
We refer to Ref.~\onlinecite{Bintrim_2021} for a detailed discussion of the up/downfolding processes of the $GW$ equations (see also Refs.~\onlinecite{Tolle_2023,Scott_2023}).

As can be readily seen in Eq.~\eqref{eq:GWlin}, the blocks $\bW^\text{2h1p}$ and $\bW^\text{2p1h}$ are coupling the 1h and 1p configuration to the 2h1p and 2p1h configurations.
Therefore, it is natural to define, within the SRG formalism, the diagonal and off-diagonal parts of the $GW$ effective Hamiltonian as
\begin{subequations}
\begin{align}
  \label{eq:diag_and_offdiag}
  \bH^\text{d}(s) &= 
  \begin{pmatrix}
    \bF	&	\bO	                	&     \bO		\\
    \bO	&	\bC^{\text{2h1p}}		&	\bO			\\
    \bO	&	\bO				&	\bC^{\text{2p1h}}	\\
  \end{pmatrix},
  \\
    \bH^\text{od}(s)  &= 
  \begin{pmatrix}
    \bO			                &     \bW^{\text{2h1p}} &     \bW^{\text{2p1h}} \\
    (\bW^{\text{2h1p}})^\dag	&	\bO	   	        &	\bO                    	\\
    (\bW^{\text{2p1h}})^\dag	&	\bO		        &	\bO	                \\
  \end{pmatrix},
\end{align}
\end{subequations}
where we omit the $s$ dependence of the matrices for the sake of brevity.
Then, our aim is to solve, order by order, the flow equation \eqref{eq:flowEquation} knowing that the initial conditions are
\begin{subequations}
\begin{align}
  \bHd{0}(0) & =   \mqty( \bF & \bO \\ \bO & \bC ),
  & 
  \bHod{0}(0) & = \bO,
  \\
  \bHd{1}(0) & = \bO, 
  &
  \bHod{1}(0) & = \mqty( \bO & \bW \\ \bW^{\dagger} & \bO ),
\end{align}
\end{subequations}
where the supermatrices 
\begin{subequations}
\begin{align}
  \bC & = \mqty( \bC^{\text{2h1p}}  &  \bO  \\  \bO  &  \bC^{\text{2p1h}}  ),
  \\
  \bW & = \mqty(  \bW^{\text{2h1p}} &  \bW^{\text{2p1h}} ),
\end{align}
\end{subequations}
collect the 2h1p and 2p1h channels.
Once the closed-form expressions of the low-order perturbative expansions are known, they can be inserted in Eq.~\eqref{eq:downfolded_sigma} to define a renormalized version of the quasiparticle equation.
In particular, we focus here on the second-order renormalized quasiparticle equation.

\subsection{Zeroth-order matrix elements}

The choice of Wegner's generator in the flow equation [see Eq.~\eqref{eq:flowEquation}] implies that the off-diagonal correction is of order $\order*{\lambda}$ while the correction to the diagonal block is at least $\order*{\lambda^2}$. \cite{Hergert_2016}
Therefore, the zeroth-order Hamiltonian is independent of $s$ and we have
\begin{equation}
  \bH^{(0)}(s) = \bH^{(0)}(0).
\end{equation}

\subsection{First-order matrix elements}

Knowing that $\bHod{0}(s)=\bO$, the first-order flow equation is
\begin{equation}
  \dv{\bH^{(1)}}{s} = \comm{\comm{\bHd{0}}{\bHod{1}}}{\bHd{0}},
\end{equation}
which gives the following system of equations
\begin{align}
  \label{eq:F0_C0}
  \dv{\bF^{(0)}}{s}&=\bO,   &  \dv{\bC^{(0)}}{s}&=\bO,
\end{align}
and
\begin{multline}
  \label{eq:W1}
  \dv{\bW^{(1)}}{s}
   = 2 \bF^{(0)}\bW^{(1)}\bC^{(0)} 
   \\
    - (\bF^{(0)})^2\bW^{(1)} - \bW^{(1)}(\bC^{(0)})^2.
\end{multline}
Equation \eqref{eq:F0_C0} implies
\begin{subequations}
\begin{align}
  \bF^{(1)}(s) &= \bF^{(1)}(0) = \bO, &
  \\
   \bC^{(1)}(s) &= \bC^{(1)}(0) = \bO,
\end{align}
\end{subequations}
and, thanks to the diagonal structure of $\bF^{(0)}$ (which is a consequence of the HF starting point) and $\bC^{(0)}$, the differential equation for the coupling block in Eq.~\eqref{eq:W1} is easily solved and yields
\begin{equation}
  W_{pq}^{\nu(1)}(s) = W_{pq}^{\nu} e^{-(\Delta_{pq}^{\nu})^2 s}.
\end{equation}
At $s=0$, $W_{pq}^{\nu(1)}(s)$ reduces to the screened two-electron integrals defined in Eq.~\eqref{eq:GW_sERI}, while,
\begin{equation}
	\lim_{s\to\infty} W_{pq}^{\nu(1)}(s) = 0.
\end{equation}
Therefore, $W_{pq}^{\nu(1)}(s)$ is a genuine renormalized two-electron screened integral.
It is worth noting the close similarity of the first-order elements with the ones derived by Evangelista in Ref.~\onlinecite{Evangelista_2014b} in the context of single- and multi-reference perturbation theory (see also Ref.~\onlinecite{Hergert_2016}).

\subsection{Second-order matrix elements}

The second-order renormalized quasiparticle equation is given by
\begin{equation}
  \label{eq:GW_renorm}
  \qty[ \widetilde{\bF}(s) + \widetilde{\bSig}(\omega = \epsilon_p; s) ] \psi_p(\bx) = \epsilon_p \psi_p(\bx),
\end{equation}
with a renormalized Fock matrix of the form
\begin{equation}
  \widetilde{\bF}(s) = \bF^{(0)}+\bF^{(2)}(s),
\end{equation}
and a renormalized dynamical self-energy
\begin{equation}
  \label{eq:srg_sigma}
  \widetilde{\bSig}(\omega; s) = \bV^{(1)}(s) \left(\omega \bI - \bC^{(0)}\right)^{-1}  (\bV^{(1)}(s))^{\dagger},
\end{equation}
with elements
\begin{equation}
  \label{eq:SRG-GW_selfenergy}
  \begin{split}
    \widetilde{\bSig}_{pq}(\omega; s) 
  &= \sum_{i\nu} \frac{W_{pi}^{\nu} W_{qi}^{\nu}}{\omega - \epsilon_i + \Omega_{\nu}} e^{-\qty[(\Delta_{pi}^{\nu})^2 + (\Delta_{qi}^{\nu})^2 ] s} \\ 
  &+ \sum_{a\nu} \frac{W_{pa}^{\nu} W_{qa}^{\nu}}{\omega - \epsilon_a - \Omega_{\nu}}e^{-\qty[(\Delta_{pa}^{\nu})^2 + (\Delta_{qa}^{\nu})^2 ] s}.
  \end{split}
\end{equation}

As can be readily seen above, $\bF^{(2)}$ is the only second-order block of the effective Hamiltonian contributing to the second-order SRG quasiparticle equation.
Collecting every second-order term in the flow equation and performing the block matrix products results in the following differential equation
\begin{multline}
  \label{eq:diffeqF2}
  \dv{\bF^{(2)}}{s}  = \bF^{(0)}\bW^{(1)}\bW^{(1),\dagger} + \bW^{(1)}\bW^{(1),\dagger}\bF^{(0)} \\ - 2 \bW^{(1)}\bC^{(0)}\bW^{(1),\dagger},
\end{multline}
which can be solved by simple integration along with the initial condition $\bF^{(2)}(0)=\bO$ to yield
\begin{multline}
  F_{pq}^{(2)}(s) 
  = \sum_{r\nu} \frac{\Delta_{pr}^{\nu}+ \Delta_{qr}^{\nu}}{(\Delta_{pr}^{\nu})^2 + (\Delta_{qr}^{\nu})^2} W_{pr}^{\nu} W_{qr}^{\nu}  
  \\
  \times \qty[1 - e^{-\qty[(\Delta_{pr}^{\nu})^2 + (\Delta_{qr}^{\nu})^2 ] s}].
\end{multline}

\begin{figure}
  \centering
  \includegraphics[width=\linewidth]{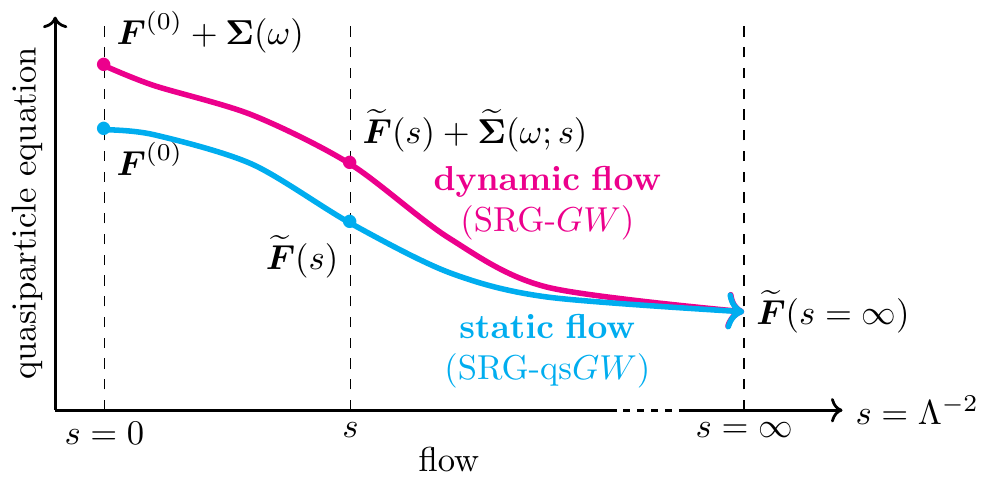}
  \caption{
    Schematic evolution of the quasiparticle equation as a function of the flow parameter $s$ in the case of the dynamic SRG-$GW$ flow (magenta) and the static SRG-qs$GW$ flow (cyan). 
    \label{fig:flow}}
\end{figure}

At $s=0$, the second-order correction vanishes, hence giving
\begin{equation}
	\lim_{s\to0} \widetilde{\bF}(s) = \bF^{(0)}.
\end{equation}
For $s\to\infty$, it tends towards the following static limit
\begin{equation}
  \label{eq:static_F2}
  \lim_{s\to\infty} \widetilde{\bF}(s) 
  = \epsilon_p \delta_{pq} 
  + \sum_{r\nu} \frac{\Delta_{pr}^{\nu}+ \Delta_{qr}^{\nu}}{(\Delta_{pr}^{\nu})^2 + (\Delta_{qr}^{\nu})^2} W_{pr}^{\nu} W_{qr}^{\nu},
\end{equation}
while the dynamic part of the self-energy [see Eq.~\eqref{eq:srg_sigma}] tends to zero, \ie,
\begin{equation}
  \lim_{s\to\infty} \widetilde{\bSig}(\omega; s) = \bO.
\end{equation}
Therefore, the SRG flow continuously transforms the dynamical self-energy $\widetilde{\bSig}(\omega; s)$ into a static correction $\widetilde{\bF}^{(2)}(s)$.
As illustrated in Fig.~\ref{fig:flow} (magenta curve), this transformation is done gradually starting from the states that have the largest denominators in Eq.~\eqref{eq:static_F2}.

For a fixed value of the energy cutoff $\Lambda$, if $\abs*{\Delta_{pr}^{\nu}} \gg \Lambda$, then $W_{pr}^{\nu} e^{-(\Delta_{pr}^{\nu})^2 s} \approx 0$, meaning that the state is decoupled from the 1h and 1p configurations, while, for $\abs*{\Delta_{pr}^{\nu}} \ll \Lambda$, we have $W_{pr}^{\nu}(s) \approx W_{pr}^{\nu}$, that is, the state remains coupled.

\begin{figure*}
  \includegraphics[width=0.8\linewidth]{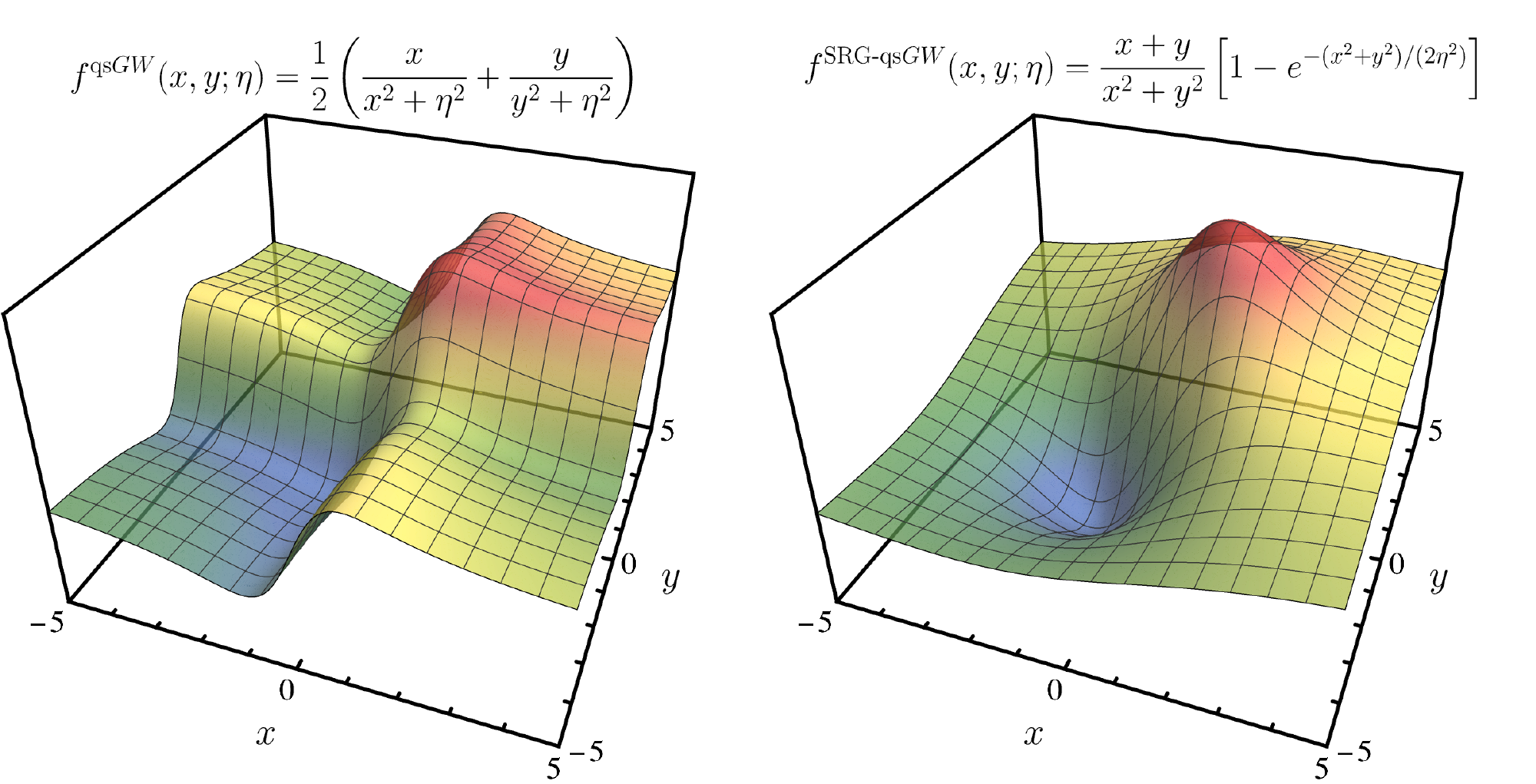}
  \caption{
    Functional form of the qs$GW$ self-energy (left) for $\eta = 1$ and the SRG-qs$GW$ self-energy (right) for $s = 1/(2\eta^2) = 1/2$.}
  \label{fig:plot}
\end{figure*}

\subsection{Alternative form of the static self-energy}

Because the large-$s$ limit of Eq.~\eqref{eq:GW_renorm} is purely static and Hermitian, the new alternative form of the self-energy reported in Eq.~\eqref{eq:static_F2} can be naturally used in qs$GW$ calculations to replace Eq.~\eqref{eq:sym_qsgw}.
Unfortunately, as we shall discuss further in Sec.~\ref{sec:results}, as $s\to\infty$, self-consistency is once again quite difficult to achieve, if not impossible.
However, one can define a more flexible new static self-energy, which will be referred to as SRG-qs$GW$ in the following, by discarding the dynamic part in Eq.~\eqref{eq:GW_renorm} (see cyan curve in Fig.~\ref{fig:flow}).
This yields a $s$-dependent static self-energy which matrix elements read
\begin{multline}
  \label{eq:SRG_qsGW}
   \Sigma_{pq}^{\SRGqsGW}(s) 
   = \sum_{r\nu} \frac{\Delta_{pr}^{\nu}+ \Delta_{qr}^{\nu}}{(\Delta_{pr}^{\nu})^2 + (\Delta_{qr}^{\nu})^2} W_{pr}^{\nu} W_{qr}^{\nu}  
   \\
   \times \qty[1 - e^{-\qty[(\Delta_{pr}^{\nu})^2 + (\Delta_{qr}^{\nu})^2 ] s} ].
\end{multline}
Note that the static SRG-qs$GW$ approximation defined in Eq.~\eqref{eq:SRG_qsGW} is straightforward to implement in existing code and is naturally Hermitian as opposed to the usual case [see Eq.~\eqref{eq:sym_qsGW}] where it is enforced by brute-force symmetrization.
Another important difference is that the SRG regularizer is energy-dependent while the imaginary shift is the same for every self-energy denominator.
Yet, these approximations are closely related because, for $\eta=0$ and $s\to\infty$, they share the same diagonal terms.

It is well-known that in traditional qs$GW$ calculations, increasing $\eta$ to ensure convergence in difficult cases is most often unavoidable.
Similarly, in SRG-qs$GW$, one might need to decrease the value of $s$ to ensure convergence.
Indeed, the fact that SRG-qs$GW$ calculations do not always converge in the large-$s$ limit is expected as, in this limit, potential intruder states have been included.
Therefore, one should use a value of $s$ large enough to include as many states as possible but small enough to avoid intruder states.

It is instructive to examine the functional form of both regularizing functions (see Fig.~\ref{fig:plot}).
These have been plotted for a regularizing parameter value of $\eta=1$, where we have set $s=1/(2\eta^2)$ such that the first-order Taylor expansion around $(x,y) = (0,0)$ of both functional forms is equal.
One can observe that the SRG-qs$GW$ surface is much smoother than its qs$GW$ counterpart.
This is due to the fact that the SRG-qs$GW$ functional at $\eta=0$, $f^{\SRGqsGW}(x,y;0)$, has fewer irregularities. 
In fact, there is a single singularity at $x=y=0$.
On the other hand, the function $f^{\qsGW}(x,y;0)$ is singular on the two entire axes, $x=0$ and $y=0$.
We believe that the smoothness of the SRG-qs$GW$ surface is the key feature that explains the faster convergence of SRG-qs$GW$ compared to qs$GW$.
The convergence properties and the accuracy of both static approximations are quantitatively gauged in Sec.~\ref{sec:results}.

To conclude this section, we briefly discussed the case of discontinuities mentioned in Sec.~\ref{sec:intro}.
Indeed, it has been previously mentioned that intruder states are responsible for both the poor convergence of qs$GW$ and discontinuities in physical quantities. \cite{Loos_2018b,Veril_2018,Loos_2020e,Berger_2021,DiSabatino_2021,Monino_2022,Scott_2023}
Is it then possible to rely on the SRG machinery to remove discontinuities?
Not directly because discontinuities are due to intruder states in the dynamic part of the quasiparticle equation.
However, as we have seen just above the functional form of the renormalized equation makes it possible to choose $s$ such that there is no intruder states in its static part.
Performing a bijective transformation of the form,
\begin{align}
  e^{- \Delta s} &= 1-e^{-\Delta t},
\end{align}
on the renormalized quasiparticle equation \eqref{eq:GW_renorm} reverses the situation and makes it possible to choose $t$ such that there is no intruder states in the dynamic part, hence removing discontinuities.
Note that, after this transformation, the form of the regularizer is actually closely related to the SRG-inspired regularizer introduced by Monino and Loos in Ref.~\onlinecite{Monino_2022}.

The intruder-state-free dynamic part of the self-energy makes it possible to define SRG-$G_0W_0$ and SRG-ev$GW$ schemes.
Although the manuscript focuses on SRG-qs$GW$, the performance of SRG-$G_0W_0$ and SRG-ev$GW$ are discussed in the {\SupInf} for the sake of completeness.
In a nutshell, the SRG regularization improves slightly the overall convergence properties of SRG-ev$GW$ without altering its performance.
Likewise, the statistical indicators for $G_0W_0$ and SRG-$G_0W_0$ are extremely close.

\section{Computational details}
\label{sec:comp_det}

Our set of systems is composed by closed-shell compounds that correspond to the 50 smallest atoms and molecules (in terms of the number of electrons) of the $GW$100 benchmark set. \cite{vanSetten_2015}
We will refer to this set as $GW$50.
Following the same philosophy as the \textsc{quest} database for neutral excited states, \cite{Loos_2020d,Veril_2021} their geometries have been optimized at the CC3/aug-cc-pVTZ basis level \cite{Christiansen_1995b,Koch_1997} using the \textsc{cfour} program. \cite{CFOUR}

The two qs$GW$ variants considered in this work have been implemented in an in-house program, named \textsc{quack}. \cite{QuAcK}
The $GW$ implementation closely follows the one of \textsc{molgw}. \cite{Bruneval_2016}
In all $GW$ calculations, we use the aug-cc-pVTZ cartesian basis set and self-consistency is performed on all (occupied and virtual) orbitals, including core orbitals.
We use (restricted) HF guess orbitals and energies for all self-consistent $GW$ calculations.
The maximum size of the DIIS space \cite{Pulay_1980,Pulay_1982} and the maximum number of iterations were set to 5 and 64, respectively.
In practice, one may achieve convergence, in some cases, by adjusting these parameters or by using an alternative mixing scheme.
However, in order to perform black-box comparisons, these parameters have been fixed to these default values.
The $\eta$ value has been set to \num{e-3} for the conventional $G_0W_0$ calculations (where we eschew linearizing the quasiparticle equation) while, for the qs$GW$ calculations, $\eta$ has been chosen as the largest value where one successfully converges the 50 systems composing the test set.

The various $GW$-based sets of values are compared with a set of reference values computed at the $\Delta$CCSD(T) level with the same basis set.
The $\Delta$CCSD(T) principal ionization potentials (IPs) and electron affinities (EAs) have been obtained using \textsc{gaussian 16} \cite{g16} (with default parameters) within the restricted and unrestricted formalism for the neutral and charged species, respectively.

\section{Results}
\label{sec:results}


\begin{figure}
  \includegraphics[width=\linewidth]{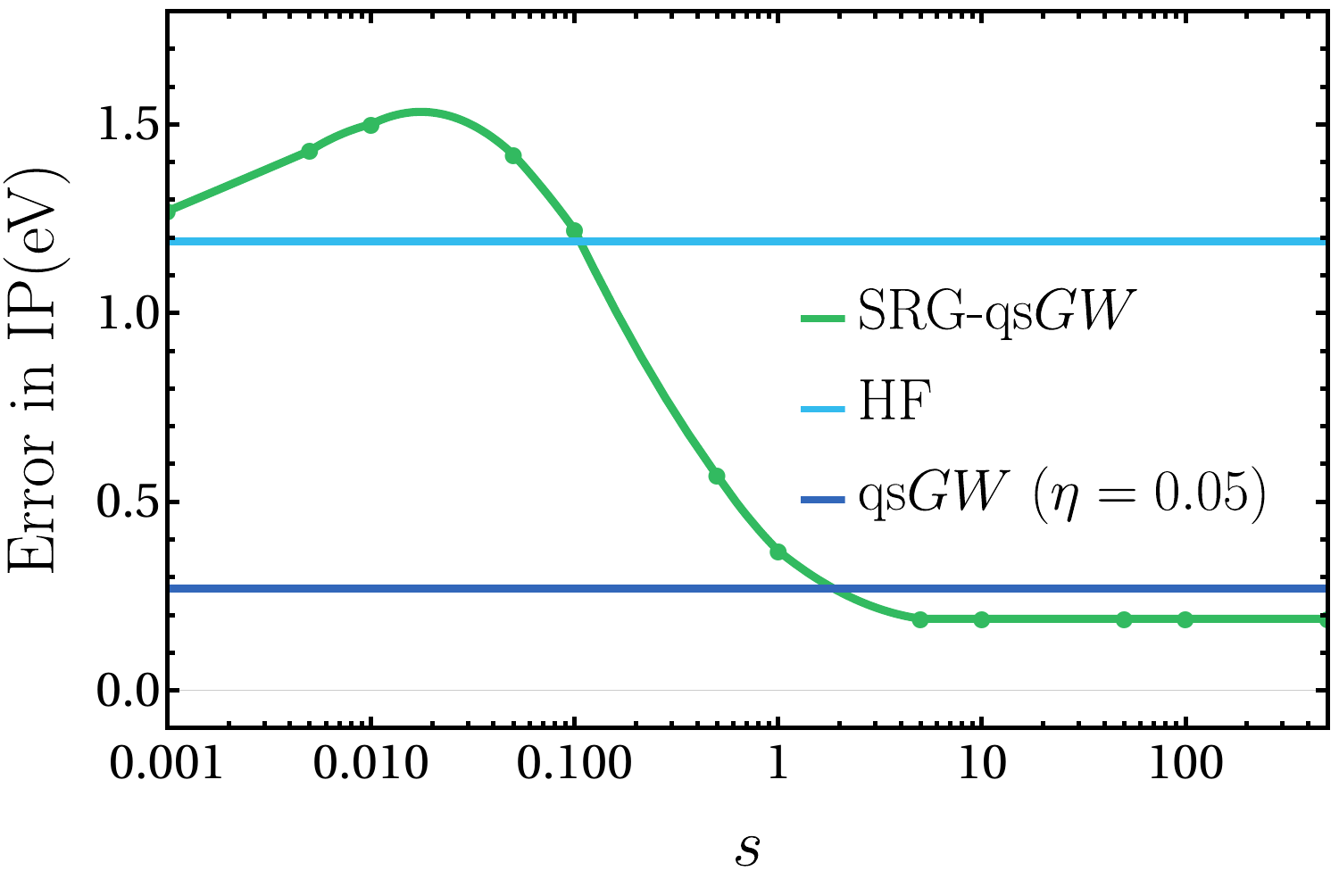}
  \caption{
    Error [with respect to $\Delta$CCSD(T)] in the principal IP of water in the aug-cc-pVTZ basis set as a function of the flow parameter $s$ for SRG-qs$GW$ (green curve).
    The HF (cyan line) and qs$GW$ (blue line) values are also reported.
    \label{fig:fig3}}
\end{figure}

\begin{figure*}
  \includegraphics[width=\linewidth]{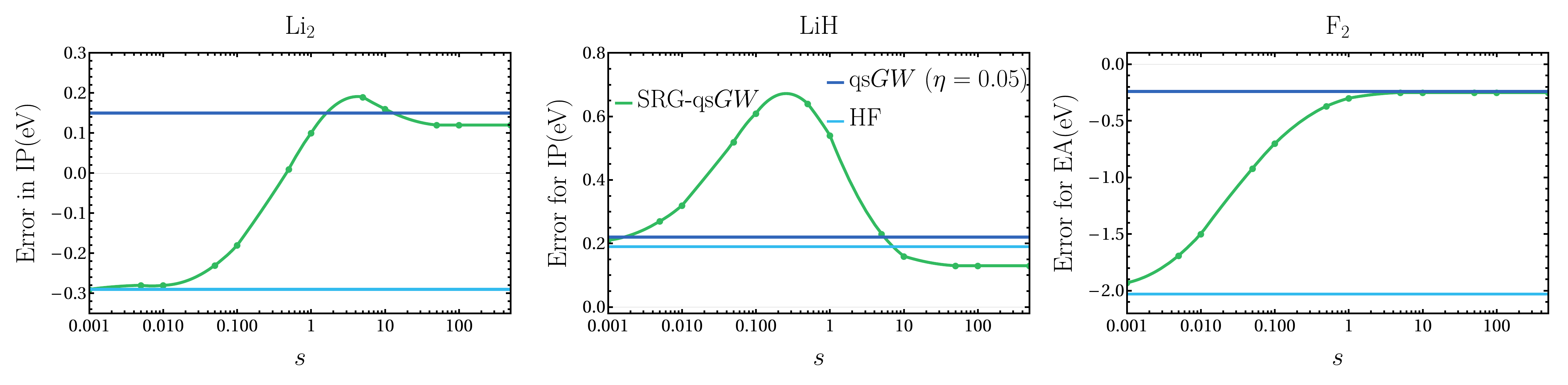}
  \caption{
    Error [with respect to $\Delta$CCSD(T)] in the principal IP of \ce{Li2}, \ce{LiH}, and the principal EA of \ce{F2} in the aug-cc-pVTZ basis set as a function of the flow parameter $s$ for the SRG-qs$GW$ method (green curves).
    The HF (cyan lines) and qs$GW$ (blue lines) values are also reported.
    \label{fig:fig4}}
\end{figure*}

\subsection{Flow parameter dependence of SRG-qs$GW$}
\label{sec:flow_param_dep}

This section starts by considering a prototypical molecular system, the water molecule, in the aug-cc-pVTZ basis set.
Figure \ref{fig:fig3} shows the error in the principal IP [with respect to the $\Delta$CCSD(T) reference value] as a function of the flow parameter in SRG-qs$GW$ (green curve). 
The corresponding HF and qs$GW$ (computed with $\eta = 0.05$) values are also reported for the sake of comparison. 
The IP at the HF level (cyan line) is too large; this is a consequence of the missing correlation and the lack of orbital relaxation in the cation, a result that is well understood. \cite{SzaboBook,Lewis_2019}
The usual qs$GW$ scheme (blue line) brings a quantitative improvement as the IP is now within \SI{0.3}{\eV} of the reference value.

At $s=0$, the SRG-qs$GW$ IP is equal to its HF counterpart as expected from the discussion of Sec.~\ref{sec:srggw}.
As $s$ grows, the IP reaches a plateau at an error that is significantly smaller than the HF starting point.
Furthermore, the value associated with this plateau is slightly more accurate than its qs$GW$ counterpart.
However, the SRG-qs$GW$ error does not decrease smoothly between the initial HF value and the large-$s$ limit. 
For small $s$, it is actually worse than the HF starting point.

This behavior as a function of $s$ can be understood by applying matrix perturbation theory to Eq.~\eqref{eq:GWlin}. \cite{Schirmer_2018}
Through second order in the coupling block, the principal IP is 
\begin{equation}
  \label{eq:2nd_order_IP}
  \text{IP} \approx - \epsilon_\text{h} - \sum_{i\nu} \frac{(W_{\text{h}}^{i\nu})^2}{\epsilon_\text{h} - \epsilon_i + \Omega_\nu} - \sum_{a\nu} \frac{(W_{\text{h}}^{a\nu})^2}{\epsilon_\text{h} - \epsilon_a - \Omega_\nu},
\end{equation}
where $\text{h}$ is the index of the highest occupied molecular orbital (HOMO).
The first term of the right-hand side of Eq.~\eqref{eq:2nd_order_IP} is the zeroth-order IP and the following two terms originate from the 2h1p and 2p1h coupling, respectively.
The denominators of the 2p1h term are positive while the denominators associated with the 2h1p term are negative.

As $s$ increases, the first states that decouple from the HOMO are the 2p1h configurations because their energy difference with respect to the HOMO is larger than the ones associated with the 2h1p block.
Therefore, for small $s$, only the last term of Eq.~\eqref{eq:2nd_order_IP} is partially included, resulting in a positive correction to the IP.
As soon as $s$ is large enough to decouple the 2h1p block, the IP starts decreasing and eventually goes below the initial value at $s=0$, as observed in Fig.~\ref{fig:fig3}.


Next, the flow parameter dependence of SRG-qs$GW$ is investigated for the principal IP of two additional molecular systems as well as the principal EA of \ce{F2}.
The left panel of Fig.~\ref{fig:fig4} shows the results for the lithium dimer, \ce{Li2}, which is an interesting case because, unlike in water, HF underestimates the reference IP.
Yet, the qs$GW$ and SRG-qs$GW$ IPs are still overestimating the reference value as in \ce{H2O}. 
Indeed, we can see that the positive increase of the SRG-qs$GW$ IP is proportionally more important than for water.
In addition, the plateau is reached for larger values of $s$ in comparison to Fig.~\ref{fig:fig3}.

Now turning to lithium hydride, \ce{LiH} (see middle panel of Fig.~\ref{fig:fig4}), we see that the qs$GW$ IP is actually worse than the fairly accurate HF value.
However, SRG-qs$GW$ does not suffer from the same problem and improves slightly the accuracy as compared to HF.

Finally, we also consider the evolution with respect to $s$ of the principal EA of \ce{F2} that is displayed in the right panel of Fig.~\ref{fig:fig4}.
The HF value is largely underestimating the $\Delta$CCSD(T) reference.
Performing a qs$GW$ calculation on top of it brings a quantitative improvement by reducing the error from \SI{-2.03}{\eV} to \SI{-0.24}{\eV}.
The SRG-qs$GW$ EA (absolute) error is monotonically decreasing from the HF value at $s=0$ to an error close to the qs$GW$ one at $s\to\infty$.



\subsection{Statistical analysis}
\label{sec:SRG_vs_Sym}

\begin{figure*}
  \includegraphics[width=\linewidth]{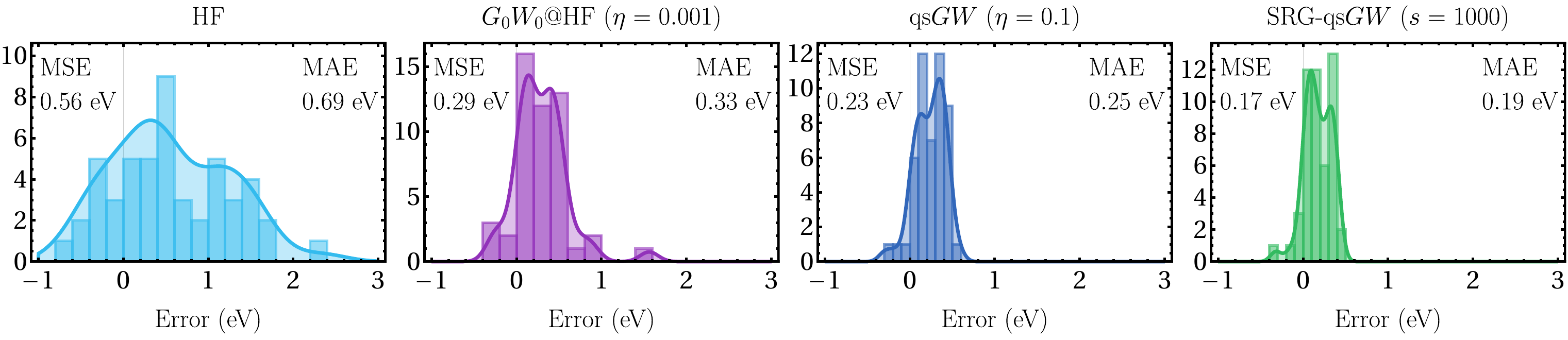}
  \caption{
    Histogram of the errors [with respect to $\Delta$CCSD(T)] for the principal IP of the $GW$50 test set calculated using HF, $G_0W_0$@HF, qs$GW$, and SRG-qs$GW$.
    All calculations are performed with the aug-cc-pVTZ basis.
    \label{fig:fig5}}
\end{figure*}

Table \ref{tab:tab1} shows the principal IP of the 50 molecules considered in this work computed at various levels of theory.
As previously mentioned, the HF approximation overestimates the IPs with a mean signed error (MSE) of \SI{0.56}{\eV} and a mean absolute error (MAE) of \SI{0.69}{\eV}.
Performing a $G_0W_0$ calculation on top of this mean-field starting point, $G_0W_0$@HF, reduces by more than a factor two the MSE and MAE, \SI{0.29}{\eV} and \SI{0.33}{\eV}, respectively.
However, there are still outliers with large errors. 
For example, the IP of \ce{N2} is overestimated by \SI{1.56}{\eV}, a large discrepancy that is due to the HF starting point.
Self-consistency mitigates the error of the outliers as the MAE at the qs$GW$ level is now \SI{0.57}{\eV} and the standard deviation of the error (SDE) is decreased from \SI{0.31}{\eV} for $G_0W_0$@HF to \SI{0.18}{\eV} for qs$GW$.
In addition, the MSE and MAE (\SI{0.23}{\eV} and \SI{0.25}{\eV}, respectively) are also slightly improved with respect to $G_0W_0$@HF.

Let us now turn to our new method, the SRG-qs$GW$ self-consistent scheme.
Table \ref{tab:tab1} shows the SRG-qs$GW$ values for $s=\num{e3}$.
The statistical descriptors corresponding to this alternative static self-energy are all improved with respect to qs$GW$.
In particular, the MSE and MAE are decreased by \SI{0.06}{\eV}.
Of course, these are small improvements but this is done with no additional computational cost and it can be easily implemented in existing code by changing the form of the static self-energy.
The evolution of the statistical descriptors with respect to the various methods considered in Table \ref{tab:tab1} is graphically illustrated in Fig.~\ref{fig:fig4}.
The decrease of the MSE and SDE correspond to a shift of the maximum of the distribution toward zero and a contraction of the distribution width, respectively.

\begin{table*}
  \caption{Principal IP and EA (in eV) of the $GW$50 test set calculated using $\Delta$CCSD(T) (reference), HF, $G_0W_0$@HF, qs$GW$, and SRG-qs$GW$. 
  The statistical descriptors associated with the errors with respect to the reference values are also reported.
  All calculations are performed with the aug-cc-pVTZ basis.}
  \label{tab:tab1}
  \begin{ruledtabular}
    \begin{tabular}{ldddddddddd}
    & \mc{5}{c}{Principal IP} & \mc{5}{c}{Principal EA} \\
    \cline{2-6} \cline{7-11}
    & \mcc{$\Delta\text{CCSD(T)}$} & \mcc{HF} & \mcc{$G_0W_0$@HF} & \mcc{qs$GW$} & \mcc{SRG-qs$GW$} & \mcc{$\Delta\text{CCSD(T)}$} & \mcc{HF} & \mcc{$G_0W_0$@HF} & \mcc{qs$GW$} & \mcc{SRG-qs$GW$} \\
    Mol.  &   \mcc{(Ref.)} &  & \mcc{($\eta=\num{e-3}$)} & \mcc{($\eta=\num{e-1}$)} & \mcc{($s=\num{e3}$)} & \mcc{(Ref.)} &  & \mcc{($\eta=\num{e-3}$)} & \mcc{($\eta=\num{e-1}$)} & \mcc{($s=\num{e3}$)} \\
    \hline
    \ce{He}     & 24.54 & 24.98 & 24.59 & 24.58 & 24.55 & -2.66 & -2.70 & -2.66 & -2.66 & -2.66 \\
    \ce{Ne}     & 21.47 & 23.15 & 21.46 & 21.83 & 21.59 & -5.09 & -5.47 & -5.25 & -5.19 & -5.19 \\
    \ce{H2}     & 16.40 & 16.16 & 16.49 & 16.45 & 16.45 & -1.35 & -1.33 & -1.28 & -1.28 & -1.28 \\
    \ce{Li2}    &  5.25 &  4.96 &  5.38 &  5.40 &  5.37 &  0.34 & -0.08 &  0.17 &  0.18 &  0.21 \\
    \ce{LiH}    &  8.02 &  8.21 &  8.22 &  8.25 &  8.15 & -0.29 &  0.20 &  0.27 &  0.27 &  0.27 \\
    \ce{HF}     & 16.15 & 17.69 & 16.25 & 16.45 & 16.34 & -0.66 & -0.81 & -0.71 & -0.70 & -0.70 \\
    \ce{Ar}     & 15.60 & 16.08 & 15.72 & 15.61 & 15.63 & -2.55 & -2.97 & -2.68 & -2.64 & -2.65 \\
    \ce{H2O}    & 12.69 & 13.88 & 12.90 & 12.98 & 12.88 & -0.61 & -0.80 & -0.68 & -0.65 & -0.66 \\
    \ce{LiF}    & 11.47 & 12.91 & 11.40 & 11.75 & 11.58 &  0.35 &  0.29 &  0.33 &  0.33 &  0.33 \\
    \ce{HCl}    & 12.67 & 12.98 & 12.78 & 12.77 & 12.72 & -0.57 & -0.79 & -0.64 & -0.63 & -0.63 \\
    \ce{BeO}    &  9.95 & 10.45 &  9.74 & 10.32 & 10.18 &  2.17 &  1.80 &  2.28 &  2.10 &  2.13 \\
    \ce{CO}     & 13.99 & 15.11 & 14.80 & 14.34 & 14.33 & -1.57 & -1.80 & -1.66 & -1.61 & -1.62 \\
    \ce{N2}     & 15.54 & 16.68 & 17.10 & 15.93 & 15.91 & -2.37 & -2.20 & -2.10 & -2.10 & -2.10 \\
    \ce{CH4}    & 14.39 & 14.83 & 14.76 & 14.67 & 14.63 & -0.65 & -0.79 & -0.70 & -0.68 & -0.68 \\
    \ce{BH3}    & 13.31 & 13.59 & 13.68 & 13.62 & 13.59 & -0.09 & -0.81 & -0.46 & -0.29 & -0.30 \\
    \ce{NH3}    & 10.91 & 11.69 & 11.22 & 11.18 & 11.10 & -0.61 & -0.80 & -0.68 & -0.66 & -0.66 \\
    \ce{BF}     & 11.15 & 11.04 & 11.34 & 11.19 & 11.18 & -0.80 & -1.06 & -0.90 & -0.87 & -0.86 \\
    \ce{BN}     & 12.05 & 11.55 & 11.76 & 11.89 & 11.90 &  3.02 &  2.97 &  3.90 &  3.41 &  3.44 \\
    \ce{SH2}    & 10.39 & 10.49 & 10.51 & 10.50 & 10.45 & -0.52 & -0.76 & -0.60 & -0.58 & -0.59 \\
    \ce{F2}     & 15.81 & 18.15 & 16.35 & 16.27 & 16.22 &  0.32 & -1.71 & -0.53 &  0.10 &  0.07 \\
    \ce{MgO}    &  7.97 &  8.75 &  8.40 &  8.54 &  8.36 &  1.54 &  1.40 &  1.64 &  1.72 &  1.71 \\
    \ce{O3}     & 12.85 & 13.29 & 13.56 & 13.34 & 13.27 &  1.82 &  1.32 &  2.19 &  2.23 &  2.17 \\
    \ce{C2H2}   & 11.45 & 11.16 & 11.57 & 11.46 & 11.43 & -0.80 & -0.80 & -0.71 & -0.71 & -0.71 \\
    \ce{HCN}    & 13.76 & 13.50 & 13.86 & 13.75 & 13.73 & -0.53 & -0.61 & -0.52 & -0.55 & -0.54 \\
    \ce{B2H6}   & 12.27 & 12.84 & 12.81 & 12.67 & 12.64 & -0.52 & -0.64 & -0.56 & -0.55 & -0.55 \\
    \ce{CH2O}   & 10.93 & 12.09 & 11.39 & 11.33 & 11.25 & -0.60 & -0.70 & -0.61 & -0.62 & -0.62 \\
    \ce{C2H4}   & 10.69 & 10.26 & 10.74 & 10.70 & 10.67 & -1.90 & -0.86 & -0.75 & -0.73 & -0.74 \\
    \ce{SiH4}   & 12.79 & 13.23 & 13.22 & 13.15 & 13.11 & -0.53 & -0.69 & -0.59 & -0.57 & -0.58 \\
    \ce{PH3}    & 10.60 & 10.60 & 10.79 & 10.76 & 10.73 & -0.51 & -0.71 & -0.58 & -0.56 & -0.57 \\
    \ce{CH4O}   & 11.09 & 12.30 & 11.55 & 11.49 & 11.39 & -0.59 & -0.76 & -0.64 & -0.62 & -0.63 \\
    \ce{H2NNH2} &  9.49 & 10.38 &  9.84 &  9.81 &  9.73 & -0.60 & -0.82 & -0.69 & -0.65 & -0.65 \\
    \ce{HOOH}   & 11.51 & 13.17 & 11.96 & 11.95 & 11.86 & -0.96 & -0.89 & -0.75 & -0.72 & -0.72 \\
    \ce{KH}     &  6.32 &  6.61 &  6.44 &  6.50 &  6.38 &  0.30 &  0.21 &  0.28 &  0.28 &  0.28 \\
    \ce{Na2}    &  4.93 &  4.53 &  4.98 &  5.03 &  5.01 &  0.36 & -0.01 &  0.26 &  0.27 &  0.30 \\
    \ce{HN3}    & 10.77 & 11.00 & 11.12 & 10.92 & 10.89 & -0.51 & -0.75 &  -0.6 & -0.56 & -0.56 \\
    \ce{CO2}    & 13.80 & 14.82 & 14.24 & 14.12 & 14.06 & -0.88 & -1.22 & -0.98 & -0.95 & -0.95 \\
    \ce{PN}     & 11.90 & 12.00 & 12.33 & 12.12 & 12.09 & -0.02 & -0.72 & -0.03 &  0.02 &  0.00 \\
    \ce{CH2O2}  & 11.54 & 12.94 & 12.00 & 11.97 & 11.88 & -0.63 & -0.79 & -0.69 & -0.66 & -0.67 \\
    \ce{C4}     & 11.43 & 11.61 & 11.77 & 11.57 & 11.54 &  2.38 &  0.58 &  2.24 &  2.29 &  2.30 \\
    \ce{C3H6}   & 10.83 & 11.25 & 11.20 & 11.07 & 11.03 & -0.94 & -0.88 & -0.75 & -0.73 & -0.73 \\
    \ce{C2H3F}  & 10.63 & 10.48 & 10.84 & 10.73 & 10.69 & -0.65 & -0.80 & -0.69 & -0.68 & -0.68 \\
    \ce{C2H4O}  & 10.29 & 11.64 & 10.84 & 10.74 & 10.66 & -0.54 & -0.69 & -0.56 & -0.57 & -0.57 \\
    \ce{C2H6O}  & 10.82 & 12.05 & 11.37 & 11.25 & 11.15 & -0.58 & -0.78 & -0.65 & -0.62 & -0.62 \\
    \ce{C3H8}   & 12.13 & 12.73 & 12.61 & 12.51 & 12.46 & -0.63 & -0.83 & -0.70 & -0.67 & -0.67 \\
    \ce{NaCl}   &  9.10 &  9.60 &  9.20 &  9.25 &  9.16 &  0.67 &  0.56 &  0.64 &  0.64 &  0.64 \\
    \ce{P2}     & 10.72 & 10.05 & 10.49 & 10.43 & 10.40 &  0.43 & -0.35 &  0.47 &  0.48 &  0.47 \\
    \ce{MgF2}   & 13.93 & 15.46 & 13.94 & 14.23 & 14.07 &  0.29 & -0.03 &  0.15 &  0.21 &  0.21 \\
    \ce{OCS}    & 11.23 & 11.44 & 11.52 & 11.37 & 11.32 & -1.43 & -1.27 & -1.03 & -0.97 & -0.98 \\
    \ce{SO2}    & 10.48 & 11.47 & 11.38 & 10.85 & 10.82 &  2.24 &  1.84 &  2.82 &  2.74 &  2.68 \\
    \ce{C2H3Cl} & 10.17 & 10.13 & 10.39 & 10.27 & 10.24 & -0.61 & -0.79 & -0.66 & -0.65 & -0.65 \\
	\hline
    MSE  &  &  0.56 &  0.29 &  0.23 &  0.17 &  & -0.25 &  0.02 &  0.04 &  0.04 \\
    MAE  &  &  0.69 &  0.33 &  0.25 &  0.19 &  &  0.31 &  0.16 &  0.13 &  0.12 \\
    RMSE &  &  0.87 &  0.43 &  0.29 &  0.23 &  &  0.49 &  0.28 &  0.23 &  0.22 \\
    SDE  &  &  0.68 &  0.31 &  0.18 &  0.16 &  &  0.43 &  0.29 &  0.23 &  0.22 \\
    Min  &  & -0.67 & -0.29 & -0.29 & -0.32 &  & -2.03 & -0.85 & -0.22 & -0.25 \\
    Max  &  &  2.34 &  1.56 &  0.57 &  0.42 &  &  1.04 &  1.15 &  1.17 &  1.16 \\
    \end{tabular}
  \end{ruledtabular}
\end{table*}

\begin{figure}
  \centering
  \includegraphics[width=\linewidth]{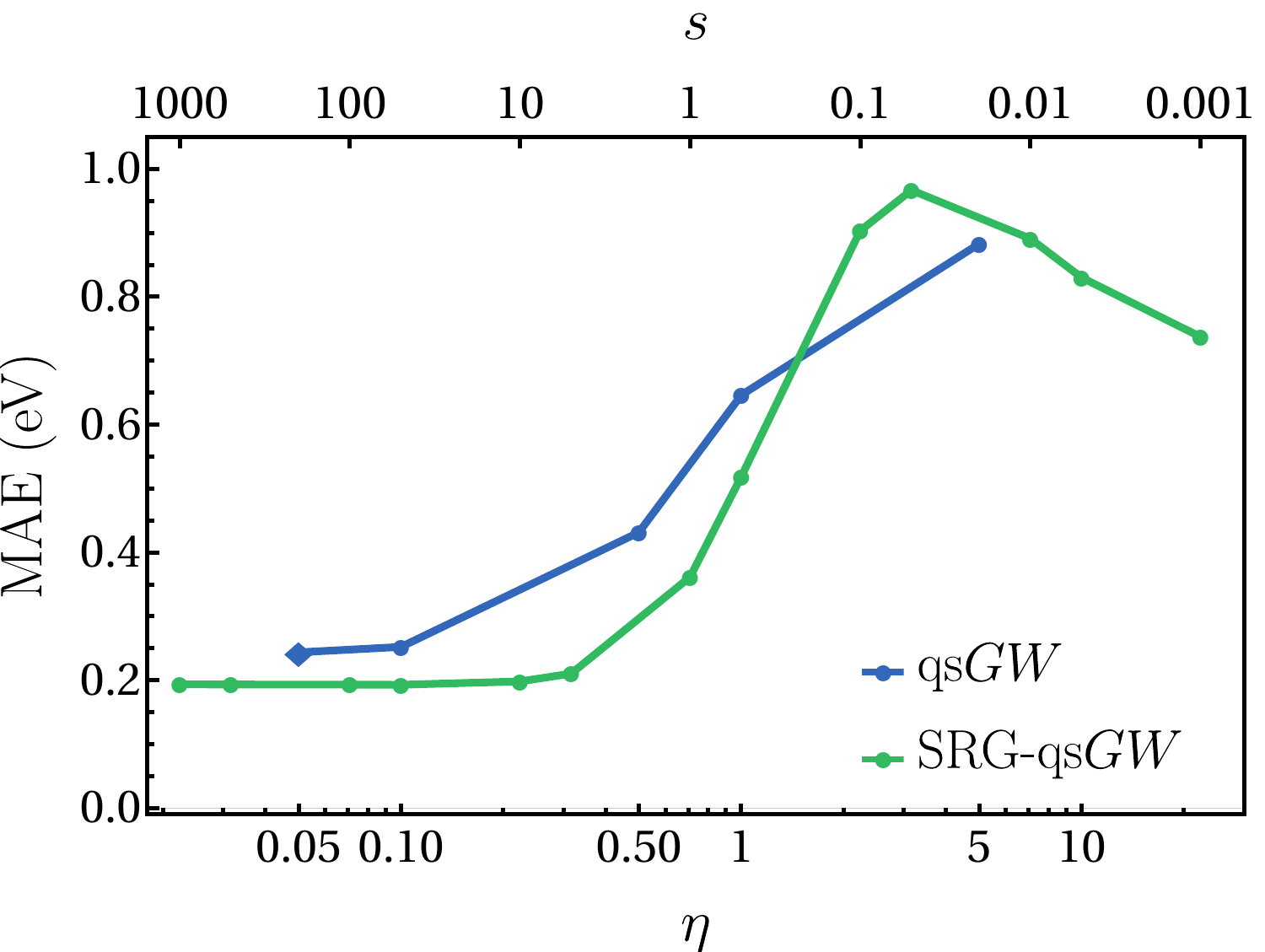}
  \caption{
    Evolution of the SRG-qs$GW$ (green) and qs$GW$ (blue) MAEs for the principal IPs of the $GW$50 test set as functions of $s$ and $\eta$, respectively. 
    The bottom and top axes are related by $s=1/(2\eta^2)$. 
    A different marker has been used for qs$GW$ at $\eta=0.05$ because the MAE includes only 48 molecules.
    \label{fig:fig6}}
\end{figure}

In addition to this improvement in terms of accuracy, the SRG-qs$GW$ scheme has been found to be much easier to converge than its qs$GW$ parent.
Indeed, up to $s=\num{e3}$, it is straightforward to reach self-consistency for the 50 compounds at the SRG-qs$GW$ level.
For $s=\num{5e3}$, convergence could not be attained for 11 systems out of 50.
However, this is not a serious issue as the MAE of the test set is already well converged at $s=\num{e3}$.
This is illustrated by the green curve of Fig.~\ref{fig:fig6} which shows the evolution of the SRG-qs$GW$ MAE with respect to $s$.
The convergence plateau of the MAE is reached around $s=50$ while the convergence problems arise for $s>\num{e3}$.
Therefore, for future studies using the SRG-qs$GW$ method, a default value of the flow parameter equal to $\num{5e2}$ or $\num{e3}$ is recommended.

On the other hand, the qs$GW$ convergence behavior is more erratic as shown by the blue curve of Fig.~\ref{fig:fig6} where we report the variation of the qs$GW$ MAE as a function of $\eta=\sqrt{1/(2s)}$.
At $\eta=\num{e-2}$ ($s=\num{5e3}$), convergence could not be reached for 13 molecules while 2 systems were already problematic at $\eta=\num{5e-2}$ ($s=200$).
These convergence problems are much more dramatic than for SRG-qs$GW$ because the MAE has not reached its limiting value before these issues arise.
For example, out of the 37 molecules that could be converged for $\eta=\num{e-2}$, the variation of the IP with respect to $\eta=\num{5e-2}$ can go up to \SI{0.1}{\eV}.

This difference in behavior is due to the energy (in)dependence of the regularizers.
The SRG regularizer first incorporates the terms with a large denominator and subsequently adds the intruder states.
Conversely, the imaginary shift regularizer treats all terms equivalently.

\begin{figure*}
  \includegraphics[width=\linewidth]{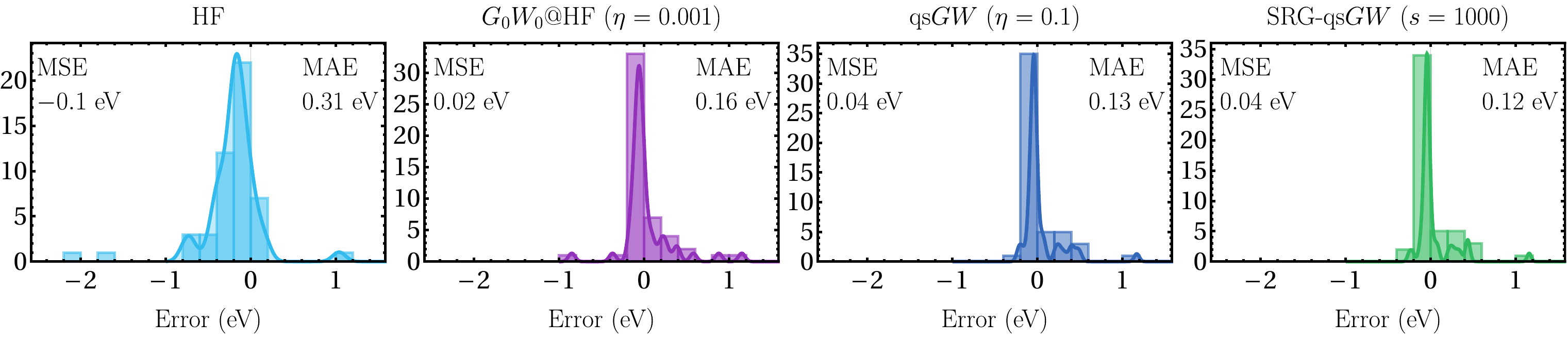}
  \caption{
    Histogram of the errors [with respect to $\Delta$CCSD(T)] for the principal EA of the $GW$50 test set calculated using HF, $G_0W_0$@HF, qs$GW$, and SRG-qs$GW$.
    All calculations are performed with the aug-cc-pVTZ basis.
   \label{fig:fig7}}
\end{figure*}

Finally, we compare the performance of HF, $G_0W_0$@HF, qs$GW$, and SRG-qs$GW$ again but for the principal EAs of $GW$50.
The raw data are reported in Table \ref{tab:tab1} while the corresponding histograms of the error distribution are plotted in Fig.~\ref{fig:fig7}.
The HF EAs are, on average, underestimated with a MAE of \SI{0.31}{\eV} and some clear outliers: \SI{-2.03}{\eV} for \ce{F2} and \SI{1.04}{\eV} for \ce{CH2O}, for example.
$G_0W_0$@HF mitigates the average error (MAE equals to \SI{0.16}{\eV}) but the minimum and maximum error values are not satisfactory.
The performance of the two qs$GW$ schemes are quite similar for EAs with MAEs of the order of \SI{0.1}{\eV}.
These two partially self-consistent methods reduce also the minimum errors but, interestingly, they do not decrease the maximum error compared to HF.

Note that a positive EA indicates a bounded anion state, which can be accurately described by the methods considered in this study. 
However, a negative EA suggests a resonance state, which is beyond the scope of the methods used in this study, including the $\Delta$CCSD(T) reference. 
As such, it is not advisable to assign a physical interpretation to these values.
Nonetheless, it is possible to compare $GW$-based and $\Delta$CCSD(T) values in such cases, provided that the comparison is limited to a given basis set.

\section{Conclusion}
\label{sec:conclusion}

The present manuscript applies the similarity renormalization group (SRG) to the $GW$ approximation of many-body perturbation theory, which is known to be plagued by intruder states.
The problems caused by intruder states in many-body perturbation theory are numerous but here we focus on the convergence issues caused by them.

SRG's central equation is the flow equation, which is usually solved numerically but can be solved analytically for low perturbation order. 
Applying this approach in the $GW$ context yields closed-form renormalized expressions for the Fock matrix elements and the screened two-electron integrals.
These renormalized quantities lead to a regularized $GW$ quasiparticle equation, referred to as SRG-$GW$, which is the main result of this work.

By isolating the static component of SRG-$GW$, we obtain an alternative Hermitian and intruder-state-free self-energy that can be used in the context of qs$GW$ calculations.
This new variant is called SRG-qs$GW$.
Additionally, we demonstrate how SRG-$GW$ can effectively resolve the discontinuity problems that arise in $GW$ due to intruder states.
This provides a first-principles justification for the SRG-inspired regularizer proposed in Ref.~\onlinecite{Monino_2022}.

We first study the flow parameter dependence of the SRG-qs$GW$ IPs for a few test cases.
The results show that the IPs gradually evolve from the HF starting point at $s=0$ to a plateau value for $s\to\infty$ that is much closer to the $\Delta$CCSD(T) reference than the HF initial value.
For small values of the flow parameter, the SRG-qs$GW$ IPs are actually worse than their starting point.
Therefore, it is advisable to use the largest possible value of $s$, similar to qs$GW$ calculations where one needs to use the smallest possible $\eta$ value.

Next, we gauge the accuracy of the SRG-qs$GW$ principal IP for a test set of 50 atoms and molecules (referred to as $GW$50).
The results show that, on average, SRG-qs$GW$ is slightly better than its qs$GW$ parent.
Despite the fact that the increase in accuracy is relatively modest, it comes with no additional computational cost and is straightforward to implement, as only the expression of the static self-energy needs to be modified.
Moreover, SRG-qs$GW$ calculations are much easier to converge than their traditional qs$GW$ counterparts thanks to the intruder-state-free nature of SRG-qs$GW$.

Finally, the principal EAs of the $GW$50 set are also investigated.
It is found that the performances of qs$GW$ and SRG-qs$GW$ are quite similar in this case.
However, it should be noted that most of the anions of the $GW$50 set are resonance states, and the associated physics cannot be accurately described by the methods considered in this study. 
Therefore, test sets of molecules with bound anions, such as this one of organic electron-acceptor molecules, \cite{Richard_2016,Gallandi_2016,Knight_2016,Dolgounitcheva_2016} and their accompanying accurate reference values are greatly valuable to the many-body perturbation theory community.

\acknowledgements{
The authors thank Francesco Evangelista for inspiring discussions.
This project has received funding from the European Research Council (ERC) under the European Union's Horizon 2020 research and innovation programme (Grant agreement No.~863481).}
\bibliography{SRGGW}

\end{document}